# Boundary Labeling in a Circular Orbit

**Annika Bonerath** ✉ ⓘ
University of Bonn, Germany

**Martin Nöllenburg** ✉ ⓘ
Algorithms and Complexity Group, TU Wien, Austria

**Soeren Terziadis** ✉ ⓘ
TU Eindhoven, The Netherlands

**Markus Wallinger** ✉ ⓘ
Algorithms and Complexity Group, TU Wien, Austria

**Jules Wulms** ✉ ⓘ
TU Eindhoven, The Netherlands

── **Abstract** ──────────────────────────────────────────────

Boundary labeling is a well-known method for displaying short textual labels for a set of point features in a figure alongside the boundary of that figure. Labels and their corresponding points are connected via crossing-free leaders. We propose *orbital boundary labeling* as a new variant of the problem, in which (i) the figure is enclosed by a circular contour and (ii) the labels are placed as disjoint circular arcs in an annulus-shaped orbit around the contour. The algorithmic objective is to compute an orbital boundary labeling with the minimum total leader length. We identify several parameters that define the corresponding problem space: two leader types (straight or orbital-radial), label size and order, presence of candidate label positions, and constraints on where a leader attaches to its label. Our results provide polynomial-time algorithms for many variants and NP-hardness for others, using a variety of geometric and combinatorial insights.

**2012 ACM Subject Classification** Human-centered computing → Visualization; Theory of computation → Computational geometry

**Funding** *Annika Bonerath*: German Research Foundation under Germany's Excellence Strategy, EXC-2070 - 390732324 - PhenoRob.
*Soeren Terziadis*: Vienna Science and Technology Fund (WWTF) [10.47379/ICT19035] and the European Union's Horizon 2020 research and innovation programme under the Marie Skłodowska-Curie Grant Agreement No 101034253
*Markus Wallinger*: Vienna Science and Technology Fund (WWTF) [10.47379/ICT19035]

# 1 Introduction

Labeling spatial data on a map is a well-studied topic in computational geometry [1, 10, 17]. Commonly the feature points are annotated with labels that display the names or additional descriptions, ensuring non-overlapping labels to guarantee full readability. The labels are placed either next to the feature points [16] (*internal* label positions), or remotely along the contour of a bounding shape such that the feature points are connected to their labels by crossing-free leaders (*external* labeling models) [5]. Often, for high feature-point densities, the external labeling model is more advantageous since the background map is not obscured by annotations. A special case of external labeling is boundary labeling [3, 4] where the labels are attached to the (mostly rectangular) boundary of the map. The interest in visualizing data on round displays, e.g., on smartwatch faces (see Figure 1) or on round displays in cockpits, is growing, as discussed by Islam et al. [14] in their recent survey; but, from a visualization perspective, it is still an under-explored topic compared to traditional rectangular displays.

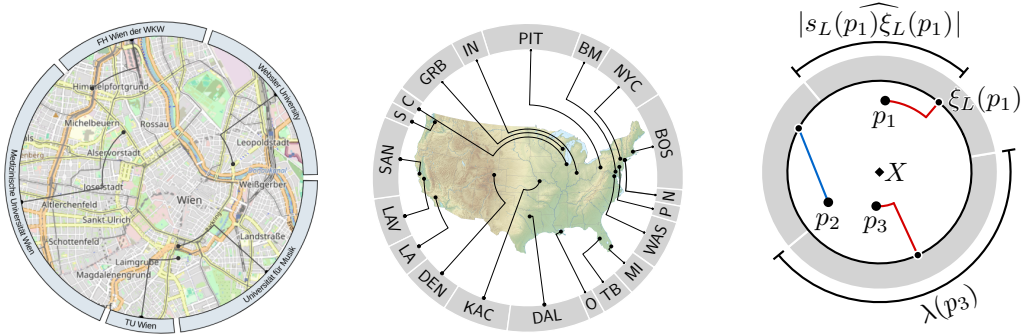

**Figure 1** An orbital labeling highlighting points of interest in a map section (left) and an orbital labeling of NFL-Championship winning teams, where the label sizes are scaled with the number of won titles (center). Our notation (right) with an SL-leader in blue and OR-leaders in red.

The design space description of Islam et al. [14] includes geospatial data representations as well as placement of text labels and icons on a round display. With this in mind, we initiate the investigation of boundary labeling for maps with circular boundaries, surrounded by a peripheral fixed-width ring, in which the labels are placed. We call these labels *orbital*. We assume that the lengths of the orbital labels are normalized, s.t., they sum up to at most the perimeter of the boundary of the map. Orbital labels can also be used for other purposes, e.g., to display donut charts representing statistical data values of different feature points within the map, such that the label sizes are proportional to the data values (Figure 1). Previous research on circular map displays considered either multirow circular labels where the sum of label lengths does not equate to the map's boundary length [12], radial labels [2,9], or horizontal labels [9,13,15]. The latter two settings are relevant on rectangular displays but not suitable for circular displays with a narrow annulus reserved for labels. Furthermore, these settings differ in their geometric properties and hence their labeling algorithms do not immediately generalize to orbital labels.

Formally, we assume that we are given a disk $D$ in the plane $\mathbb{R}^2$ centered at a point $X$. The disk contains $n$ points $P = \{p_1, \ldots, p_n\}$. We call the set $P$ of points *features* and we refer to the boundary of the disk as the *boundary* $B$. The feature closest to $X$ is denoted by $p_{\min}$. We may assume that $D$ has a radius of 1. Throughout this paper, unless otherwise specified, the *angle between* two points $p, q$ refers to the smallest angle with the center $X$ of $D$, i.e., $\min(\angle pXq, \angle qXp)$. Every feature $p \in P$ has an associated *label* representing additional information to be placed along a circular arc on the boundary starting at a point $b_1 \in B$ and ending at a point $b_2 \in B$. The circular arc along $B$ is denoted as $\widehat{b_1 b_2}$. Usually, the start and endpoint of the label are not fixed in the input, however, the length of the arc is part of the input. We represent the associated label simply as a number $\lambda(p)$, which indicates the length of the associated label. We assume that $\sum_{i=1}^n \lambda(p_i)$ is equal to the circumference of $D$, i.e., all labels can be placed in a non-overlapping way without gaps between the arcs on $B$.

In a *labeling* $L$, every feature $p \in P$ is assigned a label with starting point $s_L(p) \in B$ and an endpoint $e_L(p) \in B$, s.t., $|\widehat{s_L(p)e_L(p)}| = \lambda(p)$. We require that all labels in $L$ are pairwise non-overlapping. Additionally, every feature $p$ is connected to its label via a curve called a *leader*. We denote the length of a leader $\psi$ using $|\psi|$. We call the point on a label arc where the leader connects to $B$ the *port* $\xi_L(p)$. We represent a port by the *port ratio* $\rho(p) = \frac{|\widehat{s_L(p)\xi_L(p)}|}{\lambda(p)}$ which is the ratio of the arc from the starting point $s_L(p)$ to the port $\xi_L(p)$ and the arc from the start-point to the end-point. $\varepsilon \leq |\widehat{s_L(p)\xi_L(p)}|$ and $\varepsilon \leq |\widehat{\xi_L(p)e_L(p)}|$, i.e., there is always at least a small distance between the ends and the port of the label. A distance of $\varepsilon$ between two points on $B$ should be interpreted as a distance along $B$. Now, we define the generic orbital labeling problem.

**Table 1** An overview of the problem space and our results. Only **locked** port ratios ($A^{\lock}$) are shown. The abbreviation 'w. NP-h' denotes weakly NP-hard. Red cells represent polytime, due to the reduction to the algorithm of Benkert et al. [7]. Light red cells are covered by the reduction but are superseded by faster dedicated approaches. Blue cells with a question mark are conjectures. Cells indicate the section containing the relevant result; SL-leader results are in the full version [8].

|    |     |     | $C^{\lock}$ | | $C^{\key}$ | |
|----|-----|-----|-------------|-------------|-------------|-------------|
|    |     |     | $A_{\equiv}$ | $A_{\doteq}$ | $A_{\equiv}$ | $A_{\doteq}$ |
| OR | $O^{\lock}$ | $S_{\equiv}$ | $O(|C|n^2)$ [S. 4.1] | $O(|C|n^2)$ [S. 4.1] | $O(n^2)$ [S. 4.3] | $O(n^2)$ [S. 4.3] |
|    |             | $S_{\doteq}$ | $O(|C|n^2)$ [S. 4.1] | $O(|C|n^2)$ [S. 4.1] | $O(n^2)$ [S. 4.3] | $O(n^2)$ [S. 4.3] |
|    | $O^{\key}$  | $S_{\equiv}$ | $O(|C|^2 n)$ [S. 4.2] | $O(|C|n^3)$ [S. 3] | $O(n^5)$ [S. 3] | $O(n^5)$ [S. 3] |
|    |             | $S_{\doteq}$ | $O(|C|^4)$ [S. 3] | $O(|C|^4)$ [S. 3] | w. NP-h [S. 5] | w. NP-h [S. 5] |
| SL | $O^{\lock}$ | $S_{\equiv}$ | $O(|C|n^2)$ [S. B.1] | $O(|C|n^2)$ [S. B.1] | $O(n^2)$ ? [S. B.3] | $O(n^2)$ ? [S. B.3] |
|    |             | $S_{\doteq}$ | $O(|C|n^2)$ [S. B.1] | $O(|C|n^2)$ [S. B.1] | $O(n^2)$ ? [S. B.3] | $O(n^2)$ ? [S. B.3] |
|    | $O^{\key}$  | $S_{\equiv}$ | $O(|C|^2 n)$ [S. B.2] | | | |
|    |             | $S_{\doteq}$ | | | w. NP-h [S. B.4] | w. NP-h [S. B.4] |

▶ **Problem 1** (ORBITAL BOUNDARY LABELING). *Given a disk D, containing n feature points P with their labels $\lambda(p)$ for $p \in P$, compute a labeling L, in which all leaders are pairwise interior-disjoint and where the sum of leader lengths is minimal.*[1]

A labeling in which the sum over all leader lengths is minimum is also called a *leader length minimal labeling*. We consider two *leader types* in this paper. A *straight-line* leader or SL-*leader* is simply a straight-line segment starting at $p$ and ending at $\xi_L(p)$. Its *length* is the Euclidean distance between $p$ and $\xi_L(p)$. An *orbital-radial* leader or OR-*leader* consists of two parts: a (possibly empty) orbital circular arc with center point $X$ starting at the feature $p$ and ending at a bend point $q$, and a radial segment that connects $q$ to $\xi_L(p)$; see Figure 1. We call the line through $X$ and $\xi_L(p)$ the *supporting line* of the radial part. Note that for any pair of feature and port, there are exactly two possible OR-leaders. We call an OR-leader leaving its feature in clockwise direction a *clockwise leader* and analogously define *counter-clockwise leaders*. We will also refer to the OR-leader whose orbital part spans an angle larger than $\pi$ as the feature's *long* and to the other one as its *short* leader. For a feature $p_i$ let $r_i$ be the radius of the circle concentric with $D$ containing $p_i$. The length of the OR-leader can be expressed as a function $g: P \times [0, 2\pi] \to \mathbb{R}$ with $g(p_i, \theta) = r_i \theta + (1 - r_i)$, where $0 \le \theta < 2\pi$ describes the angle spanned by the orbital part of a leader connected to $p_i$.

▶ **Observation 2.** *For a fixed feature $p$, the function $g(p, \theta)$ is continuous and linear in $\theta$.*

Based on this problem description, we delineate the space of the possible problem variants and a suitable naming scheme for such variants in the following section.

---

[1] This definition does allow a leader to contain another feature, i.e., the endpoint of another leader.

## 2   Problem Space

In this paper, we refer to variants of ORBITAL BOUNDARY LABELING based on the six-dimensional $T$-*COSA*-ORBITAL BOUNDARY LABELING scheme introduced in this section. The first dimension, denoted by $T$, determines the leader type, whereas the other dimensions, denoted by variables $COSA$, characterize properties of the labels ($A$ encodes two dimensions). We use OR and SL as substitutes for $T$ in the $T$-*COSA* scheme. Without the $T$- prefix, we refer to both OR-*COSA* and SL-*COSA* (leader types are not mixed in a labeling). We mostly focus on OR-leaders, while still discussing which of the results extend to SL-leaders.

For the five dimensions regarding the labels, we use each letter of *COSA* to describe the variants for the respective dimension.

- [$C$] **Candidate port positions on the boundary.** If we are given a set $C$ of candidate positions on $B$, we require in any valid labeling $L$ that the set $\Xi_L$ of all ports in $L$ is a subset of $C$, we say *the port candidates are locked* (and use the symbol $C^{\lock}$) otherwise they are *free* ($C^{\free}$). For variants with $C^{\lock}$ we assume that sufficiently many, but no more than linearly many candidates are specified ($n \leq |C|$). Otherwise, $C^{\free}$ is the more reasonable choice.
- [$O$] **Order.** Consider the cyclic order of labels around $B$. If a certain label order is pre-specified, we say *the label order is locked* ($O^{\lock}$); otherwise, for the unconstrained setting, we say *the label order is free* ($O^{\free}$).
- [$S$] **Size of labels.** We distinguish the setting where $\lambda(p) = \frac{2\pi}{n}$ for all $p \in P$, in which case we say that *the label size is uniform* ($S_{\equiv}$), otherwise *the label size is non-uniform* ($S_{\not\equiv}$).
- [$A$] **Port position on labels.** We differentiate between *uniform port ratios*, where $\rho(p) = \rho(q)$ for all $p, q \in P$ ($A_{\equiv}$), and *non-uniform port ratios* ($A_{\not\equiv}$). Additionally the port ratios can be *fixed* as part of the input ($A^{\lock}$) or can be *free* to be chosen ($A^{\free}$).

To specify a $T$-*COSA* variant we substitute $C$, $O$, $S$ and $A$ with $C^{\free}/C^{\lock}$, $O^{\free}/O^{\lock}$, $S_{\equiv}/S_{\not\equiv}$ and $A_{\equiv}^{\free}/A_{\not\equiv}^{\free}/A_{\equiv}^{\lock}/A_{\not\equiv}^{\lock}$, respectively. Whenever a statement applies to all variants along a *COSA* dimension, we drop the sub- or superscript. For example, $C^{\free}O^{\lock}SA_{\equiv}^{\lock}$ refers to the variants where the leader style is either OR or SL, the port candidates are free ($C^{\free}$), the order is locked ($O^{\lock}$), the label sizes could be fixed to be uniform or they could be non-uniform ($S$) and all port ratios are fixed ($A^{\lock}$) to the same ($A_{\equiv}$) given value.

**Contributions.**   The remainder of the paper is structured as follows (see also Table 1). Our focus is OR-*COSA*$^{\lock}$ and we show how results for these variants extend to SL-*COSA*$^{\lock}$ and eventually to all *COSA* variants. Section 3 presents a reduction from OR-*COSA* to a problem called BOUNDARY LABELING [7], for which polynomial time algorithms are known. This approach is applicable to a number of OR-*COSA* variants. In Section 4 we introduce dedicated approaches for some variants, which improve the runtime of the reduction approach. In Section 5 we prove that OR-$C^{\free}O^{\free}S_{\not\equiv}A^{\lock}$ variants are weakly NP-hard. Section 6 outlines the extensions to straight-line leaders and free port ratios (details in Appendices B and C, respectively), while Section 7 provides some concluding remarks.

## 3   Reduction to Boundary Labeling

We begin by investigating the relation between ORBITAL BOUNDARY LABELING and an already established related problem called BOUNDARY LABELING. In an instance of BOUNDARY LABELING, we are given a set of points which are entirely left of a vertical line $\ell$. Additionally, we have a set of $n$ disjoint rectangular labels, which are placed to the right of $\ell$. The goal is to find a set of parallel-orthogonal (*po*) leaders, which consist of a line segment parallel to $\ell$ starting at a feature point and a second line segment orthogonal to $\ell$ connecting to a port on a label. If the label sizes are interchangeable, i.e., every point can be connected to any label, there is an $O(n^3)$ dynamic program [6,7], which minimizes an arbitrary badness function for a leader (under the assumption that the function can be computed in linear time). The dynamic program, which first splits the instance into horizontal strips, defined by points and the boundaries of the rectangles, then assigns a specific horizontal strip to the feature farthest from $\ell$ and then recurses on the two sub-instances defined by this assignment. One instance contains all features and rectangles above the assigned horizontal strip, the second everything below.

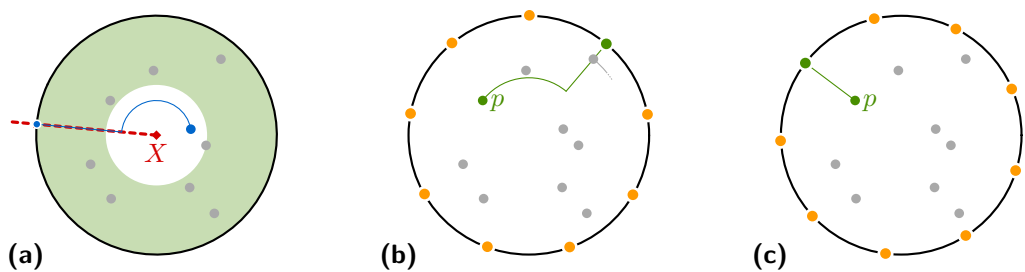

**Figure 2** All leaders of gray points in **(a)** are entirely contained within the green annulus. The radial part of the blue leader spans the entire intersection between its supporting line and the green annulus. In the labeling shown in **(b)**, a clockwise rotation of the ports is not possible without introducing a crossing between the green leader and the one in gray (indicated with a dashed line). In **(c)** any rotation clockwise or counterclockwise increases the length of the green leader. Both **(b)** and **(c)** show the implied ports in yellow.

To create an instance of BOUNDARY LABELING from an instance of ORBITAL BOUNDARY LABELING, we consider $D$ as an annulus (by introducing a small circle at the center, which does not contain any features), find a line segment $s$ orthogonal to $B$ from the boundary of the small circle to a point on $B$ and transform the annulus into a rectangle, turning $B$ into a straight line. The basis for this reduction is the following observation, which – intuitively speaking – prove that there exists a radial line, which we can use to cut and unroll our instance to remove the cyclic nature of ORBITAL BOUNDARY LABELING.

▶ **Observation 3.** *In a crossing-free leader-length minimal labeling for problems in* OR-$COSA$, *the supporting line of the radial part for the leader of $p_{min}$ intersect no other leader.*

Observation 3 is also illustrated in Figure 2a. Based on this observation we know that in any crossing free labeling, there is always a radial line, which does not intersect a leader other than the one of $p_{\min}$. Moreover it is sufficient to check all possibilities for the port of $p_{\min}$ to obtain such a line. Next, we need to show that there is only a polynomial number of possibilities for the port of $p_{\min}$. The exact number depends on the variant we are considering. If we are considering a labeling for a problem in the set OR-$C^{\bullet}OSA$, we are given the set $C$ of possible ports, leading to the following observation.

▶ **Observation 4.** *For problems in $C^{\bullet}OSA$ there are only $|C|$ possibilities for $\xi_L(p_{min})$.*

It is less obvious if and how we can discretize these options for problems without a fixed candidate set for the ports. If we are considering a problem in the set $C^{\prime}O^{\bullet}SA$, we can reduce the number of relevant options using the following lemma.

▶ **Lemma 5.** *For problems in* OR-$C^{\prime}O^{\bullet}SA$ *there are only $n^2$ possibilities for $\xi_L(p_{min})$.*

**Proof.** Let $S$ be the set of the $n$ intersection points between $B$ and any ray starting at $X$ through a feature. We prove that there is a leader-length minimal labeling, in which at least one port is in $S$. Assume that $L$ is a leader-length minimal labeling, where no port is in $S$. Consider a small rotation of all ports clockwise. If such a rotation is not possible without introducing a crossing, the radial segment of a clockwise leader already contains another feature, and therefore its port was in $S$ (see Figure 2b). Otherwise, a small enough rotation neither changes the order of labels nor does it introduce any intersections and therefore results in a new valid labeling. If this rotation decreases the total leader length, then $L$ was not optimal, which contradicts our assumption. If the total leader length stays the same, we can continue the rotation until either the orbital segment of a counter-clockwise leader reaches length 0 or the radial segment of a clockwise leader hits another feature. In both cases, its port is in $S$. If the leader length increases, we instead rotate counter-clockwise. Again, if the total leader length decreases or stays the same, the arguments above apply. Assume therefore that the total leader length again increases. Since by Observation 2 the change in leader length is linear in the angle by which we rotate, there must be a single leader that increases its length in both rotation directions, which implies that its orbital part has length 0 in $L$ (see Figure 2c).

Therefore in any optimal labeling, at least one port is in $S$. Since the order is fixed, choosing a specific label to have its port at an element in $S$ fixes all other ports and specifically the location of the port of $p_{\min}$. Therefore every element of $S$ induces $n$ possible choices for this port which results in $n^2$ choices in total. ◀

In general, the previous method does not extend to the problems in OR-$C^{\prime}O^{\prime}SA$, since choosing one element in $S$ still leaves $(n-1)!$ possible orders of the remaining labels. However, in the special case where both the label sizes and the port ratios are uniform, we know that all ports in any valid labeling are distributed equally along $B$, which implies that again every element in $S$ induces exactly $n$ possible labels (even if we do not know which label is connected to which port). Therefore, we state the following observation.

▶ **Observation 6.** *For problems in* OR-$C^{\prime}O^{\prime}S_{\equiv}A_{\equiv}$ *there are only $n^2$ possibilities for $\xi_L(p_{min})$.*

We can now proceed to describe how we create an instance of BOUNDARY LABELING based on a given instance of ORBITAL BOUNDARY LABELING.

▶ **Lemma 7.** *Given a port $\xi$ for $p_{min}$, we can reduce any problem in* OR-$CO^{\bullet}SA^{\bullet}$ *to* BOUNDARY LABELING *with po-leaders [7].*

**Proof.** We first map all features to points in the plane (also shown in Figure 3). Let $X$ be the center of $D$. For any feature $p_i \in P$, let $\alpha_i$ be the angle between $\xi$ and $p_i$ (recall that this is defined as the smaller angle formed between these two points at $X$) and let $r_i = |Xp_i|$. Using polar coordinates we then create a point $q_i = (r_i, \alpha_i)$ for every point $p_i \in P$. We now place a vertical line $\ell$ at $x = 1$. Recall that the radius of $D$ is 1 and therefore all points $q_1, \ldots, q_n$ are left of $\ell$. Since the port ratio is locked as part of the input, the fixed position of $\xi$ also fixes the exact position of the label of $p_{\min}$. If the problem is in the set OR-$CO^{\bullet}SA^{\bullet}$, the order of labels is fixed, and therefore $\xi$ also fixes the position of all other labels. For every point $p_i$ we place a rectangle of height $\lambda(p_i) - 2\varepsilon$ in the order of the labels (recall that any port has to have a minimal distance of $\varepsilon$ to the boundary of the label), s.t., the lowest point of the lowest rectangle is at height $|s_L\widehat{(p_{\min})}\xi| + \varepsilon$.

Now, the length of the radial segment of an orbital-radial leader connecting $p_i$ to a point $b \in B$ in our problem is equal to the length of the orthogonal linepart of the po-leader connecting $q_i$ to a point $v$ on $\ell$. The relation between the length of the parallel part of the po-leader and the length of the orbital segment of the OR-leader is more complicated. The length of the parallel part of the po-leader is simply the difference in $y$-coordinate between $q_i$ and $v$. Note that we mapped the clockwise angle of a point $q_i$ relative to $\xi$ to the $y$-coordinate

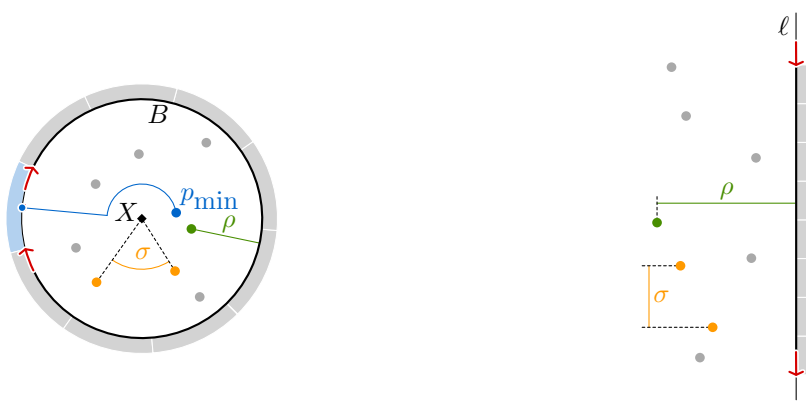

**(a)** Instance of ORBITAL BOUNDARY LABELING  **(b)** Instance of BOUNDARY LABELING

**Figure 3** An instance **(a)** of ORBITAL BOUNDARY LABELING together with a fixed port for $p_{\min}$. We use an appropriate mapping to construct an instance of BOUNDARY LABELING, whose solution **(b)** corresponds to the solution of ORBITAL BOUNDARY LABELING. This maps the angle $\sigma$ between two features in **(a)** to their horizontal distance between the mapped features in **(b)** and the distance $\rho$ of a feature to $B$ in **(a)** to the distance of the mapped feature to $\ell$ in **(b)**. Note that only the remaining non-fixed labels are part of the new instance.

of $q_i$. However, the length of the orbital part of the OR-leader is dependent on the distance of $p_i$ to $X$, i.e., two OR-leaders whose orbital part span the same angle can have different lengths. Specifically, the length of an orbital part in an OR-leader of a feature $p_i$ is exactly $r_i \cdot \alpha_i$. Therefore, we define our badness function simply as $\mathrm{bad}(q_i, v) = 1 - r_i + r_i(|\alpha_i - y(q_i)|)$, where $y(q)$ is the $y$-coordinate of $q_i$. Note that the restriction of port placement on $\ell$ to a specific range (e.g., a point in a candidate set $C$ or a point within a label corresponding to a fixed port ratio) can be encoded in this badness function too, by setting the value of $\mathrm{bad}(q_i, v) = \infty$ if $v$ lies outside of the permitted range. Finally, also note that we can check in $O(n)$ time if the split into sub-instances induced by the combination of $q$ and $v$ respects the fixed order of the labels and, if it does not, also set $\mathrm{bad}(q_i, v) = \infty$. This completes the reduction. ◂

If we do not have a fixed order of labels, we have to pay attention to the order in which the rectangles are placed. However, if the labels have uniform size, any order will result in the same set of rectangles resulting in the following observation.

▶ **Observation 8.** *Given a port $\xi$ for $p_{min}$, we can reduce any problem in* OR-$CO^{\!\!\!/}S_{\equiv}A^{\bullet}$ *to po-*BOUNDARY LABELING *[7].*

Lastly, if the label sizes are not uniform, and if a candidate set $C$ is fixed, we can adapt the dynamic program slightly to leverage this fact, by placing smaller rectangles at every possible candidate position and avoid overlap by restricting the sub-instances to appropriate ranges. The proof of the following lemma can be found in Appendix A.

▶ **Lemma 9.** *Given a port $\xi$ for $p_{min}$, any problem in* OR-$C^{\bullet}O^{\!\!\!/}S_{\equiv}A^{\bullet}$ *can be solved in $O(|C|^3)$ time by adapting the dynamic program of Benkert et al. [7].*

With all the pieces assembled, we obtain the runtime of solving a number of problem variants via this reduction. We present the relevant runtimes in the following theorem.

▶ **Theorem 10.** *For any problem $P$ in* OR-$CO^{\bullet}SA^{\bullet}$, OR-$C^{\bullet}OSA^{\bullet}$ *and* OR-$C^{\!\!\!/}O^{\!\!\!/}SA^{\bullet}$ *let $A_P$ be the size of the set of possible ports for $p_{min}$, s.t., the set can be computed in $O(A_P)$*

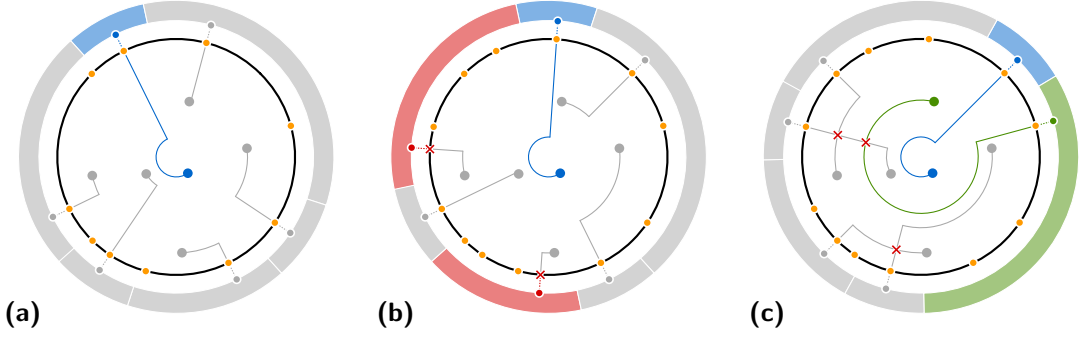

**Figure 4** Three rotations for case OR-$C^{\triangle}O^{\triangle}SA^{\triangle}$. Candidates are shown in yellow. The blue feature is the first that is placed and we iteratively test every port candidate. Due to the fixed order, the other leaders are directly obtained. In a valid labeling **(a)** all ports (obtained due to a fixed port ratio) coincide with a candidate and no two leaders cross. A labeling is invalid if ports do not coincide with a candidate – highlighted red in **(b)** – or the obtained leaders contain crossings between themselves, which is shown with the red crosses in **(c)**. In **(c)** the green leader changed from a clockwise to a counter-clockwise leader to avoid crossing the blue leader.

*time and let the time required to solve the created instance of* BOUNDARY LABELING *be* $O(B_P)$. *Then P can be solved in time* $O(A_P \cdot B_P)$.

**Proof.** We compute the set of possible ports for $p_{\min}$ as stated in Observation 4, Lemma 5 or Observation 6. For any element of these sets, we create an instance of BOUNDARY LABELING and solve it as described in Lemma 7, Observation 8 and Lemma 9. We obtain a *po*-labeling minimizing the badness function, which has a one-to-one correspondence to a leader length minimal OR-labeling of our original instance. ◀

We remark that Benkert et al. [7] point out that their algorithm can be changed to handle non-uniform labels leading to a pseudopolynomial time algorithm. By Theorem 10 this would result in pseudopolynomial time algorithms for the problems in OR-$C^{\prime}O^{\triangle}S_{\equiv}A^{\triangle}$, however, these are superseded by a dedicated approach in Section 4, so we omit further details here.

## 4 Improvement via dedicated approaches

Having established the reduction as a general approach baseline, we will now present a variety of bespoke approaches, which improve the runtime implied by Theorem 10.

### 4.1 Locked Port Candidates and Locked Order

We begin by investigating the problem set OR-$C^{\triangle}O^{\triangle}SA^{\triangle}$. Recall that we are given a set $C$ of candidate positions for the ports. The placement of the label of $p_{\min}$ determines the position of all other labels. The following lemma (which is applicable to a larger set of problems) shows that the same is true for all OR-leaders.

▶ **Lemma 11.** *For* OR-$CO^{\triangle}SA^{\triangle}$, *the choice of a port for $p_{min}$ uniquely determines the* OR-*leader of any other feature $p'$, which does not cross the* OR-*leader of $p_{min}$.*

**Proof.** Note that the orbital parts of the two possible OR-leaders connecting $p'$ to any point on $B$ form a circle concentric with $B$. Therefore one of the two orbital parts crosses the radial part of the leader of $p_{\min}$ (we also refer again to Figure 2a). The lemma follows. ◀

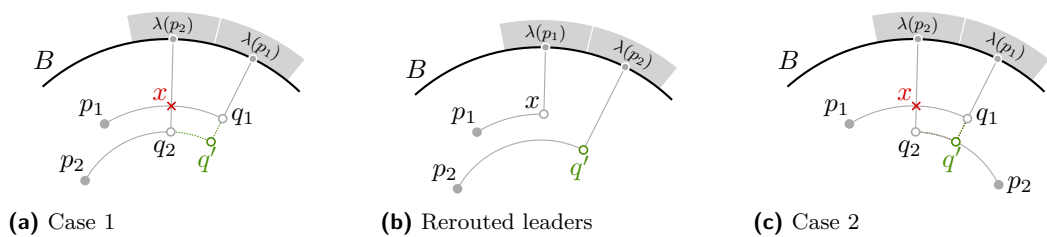

**(a)** Case 1   **(b)** Rerouted leaders   **(c)** Case 2

**Figure 5** Given a free label order $O^{\bullet}$ we can reroute the leaders to arrive at a crossing-free solution with a shorter total leader length.

By Lemma 11 it is sufficient to place $\lambda(p_{\min})$ with $\xi_L(p)$ coinciding with one candidate ($O(|C|)$ possibilities). Then we check in $O(n)$ time if the ports of the remaining labels in the correct order also coincide with candidates, in $O(n^2)$ time that no two leaders cross, and finally we compute in $O(n)$ time the total leader length, leading to a total runtime of $O(|C|n^2)$ (see Figure 4) and to the following theorem.

▶ **Theorem 12.** *The problems* OR-$C^{\bullet}O^{\bullet}SA^{\bullet}$ *can be solved in $O(|C|n^2)$ time.*

## 4.2 Locked Port Candidates, Free Order and Uniform Port Distribution

If the order can freely be chosen, then a uniform port ratio together with uniform labels guarantee that any two labels can be exchanged without creating overlap between two labels. We can thus utilize a matching algorithm to solve the problems OR-$C^{\bullet}O^{\bullet}S_{\equiv}A_{\equiv}^{\bullet}$. To obtain a crossing-free labeling we first prove the following lemma showing that a solution minimizing total leader length naturally does not contain any crossings.

▶ **Lemma 13.** *Given an instance of a problem variant in* OR-$CO^{\bullet}S_{\equiv}A_{\equiv}$ *every leader-length minimal labeling $L$ is crossing-free.*

**Proof.** Assume a leader-length minimal labeling $L$ contains two crossing leaders $\gamma_1$ and $\gamma_2$ connecting $p_1$ to its port $\xi_L(p_1) = \xi_1$ and $p_2$ to its port $\xi_L(p_2) = \xi_2$, respectively. Both leaders begin with an orbital segment $\widehat{p_1q_1}$ (or $\widehat{q_1p_1}$) and $\widehat{p_2q_2}$ (or $\widehat{q_2p_2}$), respectively, connecting to their bend points $q_1$ and $q_2$, followed by their radial straight-line segment $q_1\xi_1$ and $q_2\xi_2$. Clearly, the crossing $x$ occurs between the radial segment of the point closer to the center of $D$ and the orbital segment of the point closer to $B$. We assume, w.l.o.g., that $x = \widehat{p_1q_1} \cap q_2\xi_2$, and that $p_1$ is on the counter-clockwise end of $\widehat{p_1q_1}$. Let $q'$ be the intersection of the supporting line of $q_1\xi_1$ and the circle containing $\widehat{q_2p_2}$. There are two cases, as shown in Figure 5.

In the first case (Figure 5a), we can replace $\gamma_1$ with a curve consisting of $\widehat{p_1x}$ and $x\xi_2$ and $\gamma_2$ with curve consisting of $\widehat{p_2q'}$ and $q'\xi_1$ (Figure 5b). Since $|q_2x| = |q'q_1|$ and $|\widehat{xq_1}| > |\widehat{q_2q'}|$, the total leader length has decreased. Note that the rerouting might have introduced new crossings, but since this method reduces the total leader length, we can iteratively apply this procedure and will never obtain an already seen labeling. Since there is a finite number of possible solutions, we have to arrive at a solution, which does not contain crossings anymore (otherwise we could apply the procedure infinitely many times contradicting the finite number of possible solutions). We arrive at a labeling, that has a smaller total leader length, which is a contradiction to $L$ being optimal. While the second case (Figure 5c) looks different geometrically, we can resolve the crossing identically to the first case to again reduce the sum of leader lengths. In the special case where both features have the same distance to $X$, it is again obvious that by simply switching the ports the newly obtained leaders are a subset of the

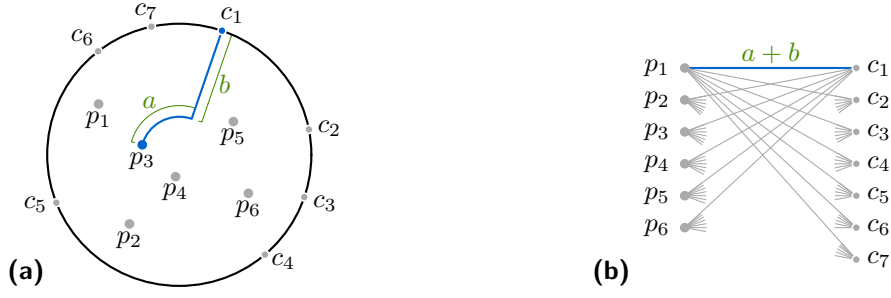

**Figure 6** Case $C^{\lock}O^{\key}S_{\equiv}A_{\equiv}^{\lock}$. Each feature $p_1, \cdots, p_6$ and port candidate $c_1, \cdots, c_7$ **(a)** introduces a vertex in the weighted complete bipartite graph **(b)**. An edge in the bipartite graph corresponds to a leader and is weighted with the leader's length.

old leaders, removing the overlap and reducing the total leader length. In all three cases, we arrived at a labeling that is better than $L$ which is a contradiction, concluding the proof. ◀

Next, we show that we can always use the shorter of the two OR-leaders.

▶ **Lemma 14.** *Given an instance of a problem variant in* OR-$CO^{\key}S_{\equiv}A_{\equiv}$ *any leader-length minimal labeling $L$ uses only the shorter of the two possible* OR-*leaders for any point.*

**Proof.** Assume $L$ contains a long OR-leader for a point $p$. Replacing the leader with the short leader between $p$ and its port yields a labeling $L'$ with a shorter total leader length. If $L'$ is crossing-free this is a contradiction to the optimality of $L$, otherwise we can iteratively apply the uncrossing procedure of the proof of Lemma 13, which, by Lemma 13, results in a crossing free labeling with the same or smaller total leader length compared to $L'$ again contradicting the optimality of $L$. ◀

With Lemmas 13 and 14 we know that every combination of feature and port defines a unique leader and the shortest possible set of such leaders is crossing free. This leads naturally to a formulation of the problem as finding a minimum-weight matching between features and ports. However, we still have to guarantee that the ports we select in such a matching are equally distributed around $B$; a requirement stemming from uniform label-sizes and fixed uniform port ratios. To this end, we state the following observation.

▶ **Observation 15.** *Given a set $C$ of candidate ports, we can partition $C$ into at most $k = \frac{|C|}{n}$ subsets $C_1, \ldots, C_k$ of size $n$, s.t., all candidates in one set $C_i$ are equally distributed around $B$. This can be done in $O(|C|^2)$ time.*

Now we can state the central result of this subsection.

▶ **Theorem 16.** *The problems in* OR-$C^{\lock}O^{\key}S_{\equiv}A_{\equiv}^{\lock}$ *can be solved in $O(|C|^2 n)$ time.*

**Proof.** We begin by creating the $k$ subsets (Observation 15). For each subset $C_i$ we let $G_i$ be a weighted complete bipartite graph between the set of features $P$ and $C_i$ using the length of the (short) leader between a feature $p \in P$ and a potential port $c \in C$ as the weight of the edge $(p, c)$; see Figure 6. For OR-leaders it is by Lemma 14 sufficient to use the length of the shorter of the two leaders. Now a minimum weight bipartite matching in $G$ corresponds to a leader-length minimal labeling. We know by Lemma 13, that such a labeling is crossing-free and an optimal solution (for $C_i$). Such a matching in a bipartite graph with $|V|$ vertices and $|E|$ edges can be computed in $O(|V|^2 \log |V| + |V||E|)$ time [11]. In our case $|V| = 2n$ and $|E| = n^2$. Therefore the runtime for one subset is $O(n^3)$ and since we run this algorithm for at most $\frac{|C|}{n}$ subsets, we arrive at a final runtime of $O(|C|^2 + |C|n^2)$. ◀

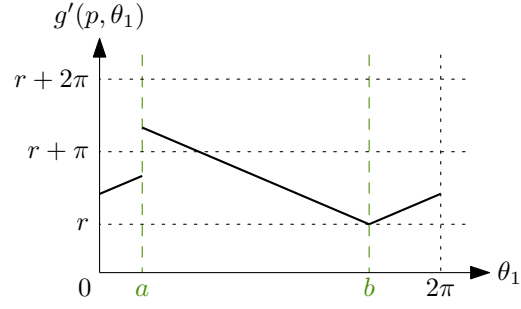

**Figure 7** An example of the function $g'(p, \theta_1)$ for a point $p$ with a distance of $1 - r$ to $X$. the points $a$ and $b$ are defined as in the proof of Lemma 17.

It is important to note that the total runtime of all iterations of the matching algorithm is strictly better than the runtime of $O(|C|n^3)$ yielded by Theorem 10, due to the preprocessing runtime of $O(|C|^2)$. Theorem 16 only guarantees an improvement for $|C| \in o(n^3)$.

## 4.3 Free Candidates and Locked Order

Now we turn to the problem set OR-$C^{\mathrm{P}}O^{\mathrm{A}}SA^{\mathrm{A}}$. Intuitively, these are problem variants, in which we can rotate the ports continuously around $B$ to obtain other labelings as long as we do not change the label order or introduce any crossings. To solve these problems, we formulate a univariate piece-wise linear function of bounded complexity, whose global minimum corresponds to an optimal labeling.

Let $\omega(i)$ be the index of the label of $p_i$ in the fixed order (assuming $\omega(1) = 1$). For any feature $p_i$ let $\theta_i^L$ be the angle between $p_i$ and $\xi_L(p_i)$ in a specific labeling $L$. Recall that by Lemma 11 the choice of the port for the innermost feature in $L$ also fixes all other ports and leaders. We will therefore replace $\theta_i^L$ with $\Theta_i(\theta_1^L)$, i.e., a function which simply returns the value of $\theta_i^L$ implied by the value of $\theta_1^L$. Since the order of labels is fixed, it is not guaranteed that every feature is connected to its port using the short OR-leader. To do so we define a Boolean variable $cw(i, \theta_1^L)$. Let $s_i$ be the intersection of $B$ and a ray starting at $X$ through $p_i$. Then $cw(i, \theta_1^L)$ is true if (in the labeling $L$ implied by the value $\theta_1$) starting at $s_i$ and traversing $B$ clockwise we encounter $\xi_L(p_i)$ before $\xi_L(p_1)$ and false otherwise.

With this we define the function $g' : P \times [0, 2\pi] \to \mathbb{R}$ as

$$g'(p_i, \theta_1) = \begin{cases} g(p_i, \Theta_i(\theta_1)) & \text{if } cw(i, \theta_1) \text{ or } i = 1 \\ g(p_i, 2\pi - \Theta_i(\theta_1)) & \text{otherwise} \end{cases} \quad (1)$$

We now provide a lemma bounding the complexity of these functions.

▶ **Lemma 17.** *Any function $g'(p_i, \theta_1)$ consists of at most three continuous linear parts in the interval $0 \le \theta_1 < 2\pi$.*

**Proof.** Assume $cw(i, 0)$ is true, i.e., we encounter $\xi_L(p_i)$ before $\xi_L(p_1)$ when traversing $B$ clockwise starting at $s_i$. While increasing $\theta_1$ the value of $cw(i, \theta_1)$ changes to false exactly when $\xi_L(p_1) = s_i$ and back to true exactly when $\xi_{L'}(p_i) = s_i$, where $L$ and $L'$ are labelings implied by some values $\theta_1 = a$ and $\theta_1 = b$, respectively, s.t., $a < b$ (see also Figure 7). Since for $\theta_1[0, a) \cup [b, 2\pi)$ we have $g'(p_i, \theta_1) = g(p_i, \Theta_i(\theta_1))$ and for $\theta_1 \in [a, b)$ we have $g'(p_i, \theta_1) = g(p_i, 2\pi - \Theta_i(\theta_1))$ and by Observation 2 the function $g$ is continuous and linear in these intervals. The case for $cw(i, 0)$ being false is symmetrical and the lemma follows. ◀

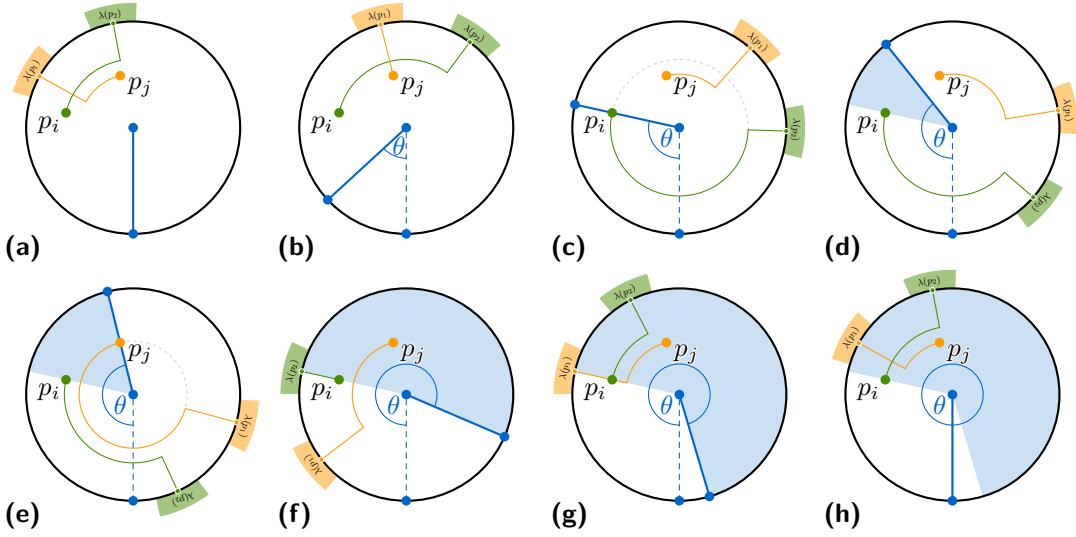

**Figure 8** Illustration of the interval (blue area) of $\theta$ in which the two leaders of the points $p_i$ and $p_j$ do not cross, while increasing $\theta$ from 0 **(a)** to $2\pi$ **(h)**. The heavy blue line is the supporting line for the radial part of the leader of $p_{\min}$. The endpoints are the value of $\theta$ for which the leader of $p_i$ changes from clockwise to counter-clockwise to avoid crossing the leader of the innermost point **(c)** and the value for which the radial part of the leader of $p_j$ crosses $p_i$ **(g)**.

We can now create a function $h(\theta_1) = \sum_{i=1}^{n} g'(p_i, \theta_1)$, which exactly captures the total leader length of a labeling implied by $\theta_1$, which adheres to the order $\omega$. However, it is important to observe that, while such a labeling does not contain any crossing between the leader of $p_1$ and any other leader, crossings between any other pair of leaders are still possible. To avoid these, we restrict $\theta_1$ to $O(n^2)$ intervals which capture exactly the values, in which the labeling implied by $\theta_1$ does not contain any crossing. We define a new Boolean variable $cr(i, j, \theta_1)$, which is true if the two OR-leaders of $p_i$ and $p_j$ that are implied by $\theta_1$, cross. Let $p_i$ be the feature closer to the center of $D$ than $p_j$. Similar to the proof of Lemma 17, we can find two values $\theta_a, \theta_b$, s.t., $cr(i, j, x) = cr(i, j, x')$ for any value $x, x' \in [\theta_a, \theta_b)$ and $cr(i, j, x) = cr(i, j, x')$ for any value $x, x' \in [0, \theta_a) \cup [\theta_b, 2\pi)$.

▶ **Lemma 18.** *Let $\mathcal{I}$ be the set of intervals, s.t., $\theta_1$ implies a labeling in which all leaders are crossing-free if and only if $\theta_1 \in I$ for some $I \in \mathcal{I}$. Then $|\mathcal{I}| \leq n^2$.*

**Proof.** For any pair of features, there is exactly one interval in which their leaders do not cross and one interval in which they do (considering the intervals modulo $2\pi$). An example is shown in Figure 8. We prove the lemma by induction. The base case is a single feature, which is always crossing free, therefore $|\mathcal{I}| = 1^2$. Assume that there are at most $n^2$ such intervals for $n$ features. We add the $(n+1)$-th feature $p$. For any of the existing $n$ features, $p$ defines a single interval in which their leaders do not cross. Consider the possibilities for adding one of these intervals $I'$ to $\mathcal{I}$. Let $I \in \mathcal{I}$. If $I \cap I' = \emptyset$, we remove $I$, if $I \cap I' = I$, we retain $I$ and if $I \cap I' = I'$, we remove $I$ and add $I'$. Otherwise, $I$ contains one or two endpoints of $I'$ but not the entire interval. If $I$ contains only one endpoint of $I'$, we remove $I$ and add $I \cap I'$. Since $I'$ is an interval, this can happen with at most two other intervals. Both times we remove one interval and add a new one, thereby not changing the size of $\mathcal{I}$. Finally, if $I$ contains both endpoints of $I'$ but not the entire interval – intuitively the intervals wrap around $B$ in two different directions – then we remove $I$ and add the two continuous parts of $I \cap I'$ increasing $|\mathcal{I}|$ by one. In this case all other intervals of $\mathcal{I}$ are either entirely contained in $I$ or disjoint

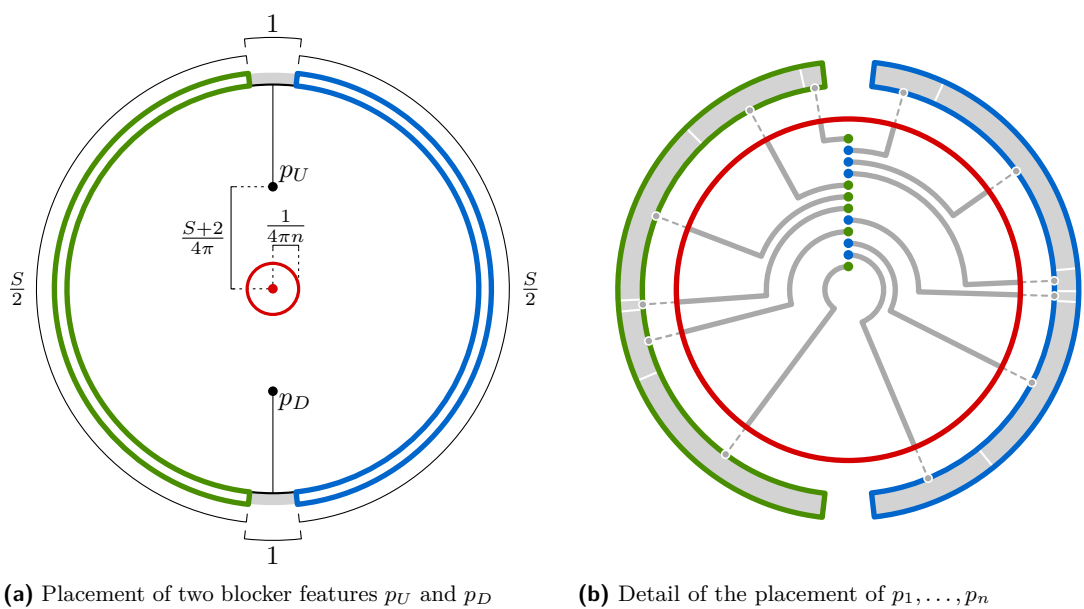

**(a)** Placement of two blocker features $p_U$ and $p_D$

**(b)** Detail of the placement of $p_1, \ldots, p_n$

**Figure 9** Visualization of the reduction. The two points $p_U$ and $p_D$ are placed close to the boundary **(a)** and all points $p_1, \ldots, p_n$ are placed in the very small red circle. A zoomed-in picture of the red circle is shown in **(b)**.

from $I$ and therefore $|\mathcal{I}|$ does not increase any further. We iteratively add all $n$ new intervals and increase $|\mathcal{I}|$ by a total of at most $n$. Therefore $|\mathcal{I}| \leq n^2 + n < n^2 + 2n + 1 = (n+1)^2$. ◂

With this, we have everything in place to prove the following theorem.

▶ **Theorem 19.** *The problem variants in* OR-$C^{\bullet}O^{\bullet}SA$ *can be solved in* $O(n^2)$ *time.*

**Proof.** We set up the function $g'(p_i, \theta_1)$ as in Equation 1 and let $h(\theta_1) = \sum_{i=1}^n g'(p_i, \theta_1)$ as above. Then we restrict $\theta_1$ to the set $\mathcal{I}$ of $n^2$ intervals (Lemma 18) in which the implied labelings do not contain any crossings. Since by Lemma 17 there are at most $O(1)$ continuous linear pieces for any function $g'$, we conclude that $h$ consists of $O(n)$ continuous linear pieces and thus also has at most $O(n)$ local minima. To find the global minimum of $h$ it is sufficient to check all $O(n)$ local minima as well as the $O(n^2)$ endpoints of the intervals in $\mathcal{I}$. ◂

## 5 Free Candidates, Free Order and Non-uniform labels are NP-hard

The results presented so far cover most of the top half of Table 1. It remains to address the problems in OR-$C^{\bullet}O^{\bullet}S_{\equiv}A^{\bullet}$. Due to the flexibility of non-uniform labels, as well as the free order and lack of a discrete candidate set, these problems turn out to be weakly NP-hard.

▶ **Theorem 20.** *Given an instance of* OR-$C^{\bullet}O^{\bullet}S_{\equiv}A^{\bullet}$ *together with $k \in \mathbb{R}$ it is (weakly)* NP-*hard to decide whether there exists a labeling $L$ with a total leader length of less than $k$.*

**Proof.** For the purpose of this proof, we will relax the requirement that $D$ has a radius of 1. The final construction can be scaled down to meet that requirement. The reduction (visualized in Figure 9) is from the weakly NP-hard problem PARTITION, where we are given a set $\mathcal{X}$ of $n$ integers with $S = \sum_{x \in \mathcal{X}} x$ and need to decide if $\mathcal{X}$ can be partitioned into two sets $\mathcal{X}_1$ and $\mathcal{X}_2$, s.t., $\sum_{x \in \mathcal{X}_1} x = \sum_{x \in \mathcal{X}_2} x = S/2$. For the reduction we place for every $x_i \in \mathcal{X}$ a feature $p_i = \left(0, \frac{i}{4\pi n^2}\right)$. Additionally we place two features $p_U = (0, r)$ and $p_D = (0, -r)$, where $r > \frac{S+2}{4\pi}$. We define $\lambda(p_i) = x_i$ for all $1 \leq i \leq n$ and $\lambda(p_U) = \lambda(p_D) = 1$. Note that $\sum_{i=1}^n \lambda(p_i) + \lambda(p_U) + \lambda(p_D) = S + 2$ and the radius of the enclosing disk is therefore $\frac{S+2}{2\pi}$.

Any feature $p_i$, s.t., $1 \leq i \leq n$ is contained in a disk of radius $\frac{1}{4\pi n}$ and circumference $\frac{1}{2n}$ centered at the origin. Let $o(i)$ and $r(i)$ be the orbital and radial part of $\gamma_L^{p_i}$, respectively. Note that the sum over all $r(i)$ is equal in all labelings. Let this sum be equal to $L_{\text{radial}}$. Further note that for any $p_i$, $o(i) < \frac{1}{2n}$. Therefore the sum $\sum_{i=1}^n o(i) < \frac{1}{2}$.

For the problem variants OR-$C^{\prime}O^{\prime}S_{\equiv}A_{\equiv}$, in any labeling $L$ the port ratios $\rho(p_U)$, and $\rho(p_D)$ are necessarily equal. For the variant OR-$C^{\prime}O^{\prime}S_{\equiv}A^{\triangleq}$ port ratios are described as part of the input and we define them, s.t., $\rho(p_U) = \rho(p_D)$. Finally we set $k = 1/2 + L_{\text{radial}}$.

If there exists a partition of $\mathcal{X}$ into two sets $\mathcal{X}_1, \mathcal{X}_2$, s.t., $\sum_{x \in \mathcal{X}_1} x = \sum_{x \in \mathcal{X}_2} x$, then we can make three observations. First, there exists a labeling $L$ in which the length of the orbital part of $\gamma_L^{p_U}$ and $\gamma_L^{p_D}$ is equal to 0 and therefore $\gamma_L^{p_U}$ and $\gamma_L^{p_D}$ are straight lines. Second, in $L$ both spaces between the labels of $p_U$ and $p_D$ are equally spaced, i.e., $|e_L(\widehat{p_U)s_L}(p_D)| = |e_L(\widehat{p_D)s_L}(p_U)|$, since $\rho(p_U) = \rho(p_D)$. Third, in a labeling $L'$, in which the length of the orbital part of $\gamma_L^{p_U}$ or $\gamma_L^{p_D}$ is not equal to 0, the sum of the length of the orbital parts of $\gamma_{L'}^{p_U}$ and $\gamma_{L'}^{p_D}$ (and therefore the sum over the lengths of all orbital parts of leaders in $L'$) is at least $\frac{2\pi}{S+2} \cdot \frac{S+2}{4\pi} = \frac{1}{2}$. This is because the difference between $e_{L'}(\widehat{p_U)s_{L'}}(p_D)$ and $e_{L'}(\widehat{p_D)s_{L'}}(p_U)$ is at least 1 (since the label sizes are integers). Therefore the sum over all leader lengths in $L$ is less than $1/2 + L_{\text{radial}}$, while in $L'$ it is at least $1/2 + L_{\text{radial}}$ and $L'$ can never be optimal.

Assume now that $\mathcal{X}$ can be partitioned into two subsets $\mathcal{X}_1, \mathcal{X}_2$, s.t., $\sum_{x \in \mathcal{X}_1} x = \sum_{x \in \mathcal{X}_2} x$. Then the labels can be equally partitioned and the leaders of $p_U$ and $p_D$ can be straight lines. Therefore the total sum of leader lengths is less than $k$. Conversely, assume no such partition exists. Then the leaders of $p_U$ and $p_D$ must together contain orbital segments of length at least $1/2$ and the total sum of leader lengths is at least $k$, concluding the proof. ◀

## 6  Extensions to SL-Leaders and Free Port Ratios

All results so far considered problem variants in OR-$COSA^{\triangleq}$, i.e., the settings which use OR-leaders and have a fixed port ratio. A number of results can be translated, sometimes with small to medium effort adaptions, to the settings using SL-leaders and/or free port ratios. In this section, we will give a high-level overview of which results can be adapted and how. The set of results for free port ratios is shown in Table 2; results for SL-leaders are included as the bottom part of both Table 1 and 2.

Some results (e.g., Theorem 12) are independent of the leader length (beyond computation of leader length). The idea of using a matching algorithm as stated in Theorem 16 also extends to the SL variant (although it requires proving that here also leader-length minimal labelings are crossing-free). But it is unclear how to find the minimum of a sum of functions describing the length of single leaders, if ports can rotate around $B$, since the number of minima such a function could have is not obvious. We include a conjecture for the settings SL-$C^{\prime}O^{\triangleq}SA^{\triangleq}$ one of which extends to a setting with free port ratio due to the equivalence of labelings. The construction of our NP-hardness reduction can in fact be reused as is for the SL-leader case, however, it requires new arguments that it still works as intended. Detailed results for SL-leaders are presented in Section B in the appendix.

While OR-leaders are related to *po*-leaders, no such immediate relation exists for SL-leaders, one reason being that OR-leaders monotonically increase their distance to $X$, which is not true for SL-leaders. Therefore the reduction to BOUNDARY LABELING explained in Section 3 does not extend to SL-leaders. It also does not work as is for free port ratios, since

▪ **Table 2** A tabular overview of the problem space and our results. Only **free** port ratios are shown. The abbreviation 'w. NP-h' stands for weakly NP-hard. Blue cells are conjectures. These results can be found in the full version [8]; cells indicate the section containing the relevant result.

| | | | $C^{🔒}$ | | $C^{🗝}$ | |
| --- | --- | --- | --- | --- | --- | --- |
| | | | $A_{\equiv}$ | $A_{\not\equiv}$ | $A_{\equiv}$ | $A_{\not\equiv}$ |
| OR | $O^{🔒}$ | $S_{\equiv}$ | $O(|C|n^2)$ [S. C.1] | $O(|C|^2n^3)$ [S. C.2] | $O(n^2)$ [S. C.1] | $O(n^6)$ [S. C.2] |
| | | $S_{\not\equiv}$ | | $O(|C|^5 n)$ [S. C.2] | | $O(n^7)$ [S. C.2] |
| | $O^{🗝}$ | $S_{\equiv}$ | $O(|C|^2n)$ [S. C.1] | $O(|C|^2n^3)$ [S. C.2] | $O(n^5)$ [S. C.1] | $O(n^6)$ [S. C.2] |
| | | $S_{\not\equiv}$ | | | w. NP-h [S. C.3] | |
| SL | $O^{🔒}$ | $S_{\equiv}$ | $O(|C|n^2)$ [S. C.1] | | $O(n^2)$ ? [S. C.1] | |
| | | $S_{\not\equiv}$ | | | | |
| | $O^{🗝}$ | $S_{\equiv}$ | $O(|C|^2n)$ [S. C.1] | | | |
| | | $S_{\not\equiv}$ | | | w. NP-h [S. C.3] | |

we relied on the fact that a position of the port of $p_{\min}$ fixes the position of its label, which is not true for free port ratios. For some instances, we can leverage candidate positions or a fixed order to prove again that only a certain set of linear or quadratic size needs to be considered for such a label. Therefore we obtain similar runtimes with an additional linear or quadratic factor. However this reduction only works for non-uniform port ratios, since otherwise the subproblems in the dynamic program of Benkert et al. [7] are not independent. Detailed results for free port ratios can be found in Section C in the appendix.

## 7 Conclusion

We have introduced orbital labeling as a variant of boundary labeling for circular boundaries, in which labels are placed as circular arcs in an annulus along the boundary. We provided a broad overview of problem variants, based on five different parameters of this labeling problem. We showed that from an algorithmic point of view, the different parameter combinations lead to distinctively different computational problems. In general, it appears that (unsurprisingly) the non-uniform label setting is computationally harder than the uniform setting. Similarly computing the layout for a fixed order is easier than for a free order of labels. The fixed candidate setting discretizes the problem and allows for an exhaustive search through all possible solutions, in contrast to the free candidate setting. In opposition to leader length, non-uniform port ratios seem to make the problem more approachable if the port ratios are also free, since free but uniform port ratios introduce a property, which has to be fulfilled globally, while free non-uniform port ratios can be fixed locally. Concerning leader types, the linear behaviour of the length of an OR-leader relative to the angle of their port with the $x$-axis allows for some approaches, which are not (or at least not immediately) applicable to SL-leaders.

It would be interesting to further investigate orbital labeling through a visualization lens as well: the variants have distinct visual styles and portray varying levels of visual complexity and uniformity. We are conducting user experiments to determine whether certain variants are superior to others.

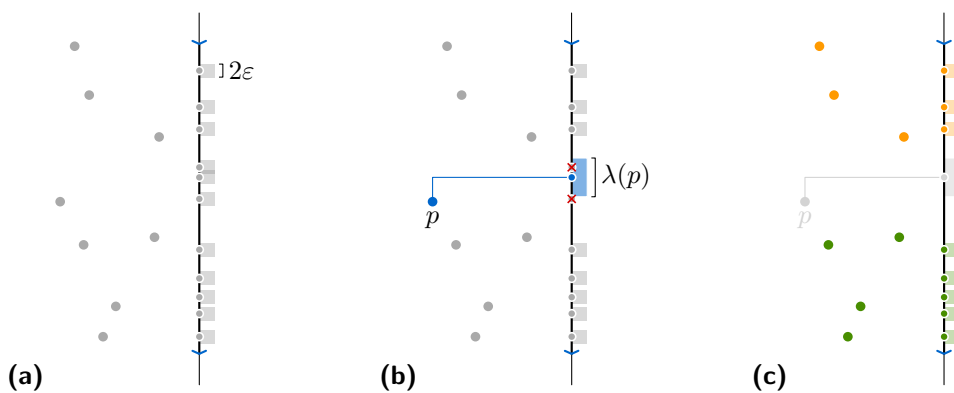

**Figure 10** For a problem variant with a given candidate set, we place small rectangles of height $2\varepsilon$ at every possible candidate point (a). If a label for a feature $p$ is placed (b) two sub-instances are created – the yellow and green sets in (c) – which exclude candidate ports, too close to the chosen port of $p$, i.e., the crossed out candidates in (b).

## A  Proof of Lemma 9

▶ **Lemma 9.** *Given a port $\xi$ for $p_{min}$, any problem in OR-$C^\triangleq O^\prime S_\pm A^\triangleq$ can be solved in $O(|C|^3)$ time by adapting the dynamic program of Benkert et al. [7].*

**Proof.** We use the same reduction as before. However, instead of placing rectangles whose size correspond to the required size for a label, we instead place rectangles of height $2\varepsilon$, s.t., any candidate $c \in C$ is exactly on the middle $y$-coordinate of one such rectangle (Figure 10a). The runtime of the algorithm scales with the number of placed labels and therefore increases from $n^3$ to $|C|^3$. The algorithm of Benkert et al. [7] repeatedly assigns a point on $\ell_v$ to the left-most point, which splits the instance into two smaller instances. Benkert et al. define a subproblem for their dynamic problem using non-overlapping rectangles that are large enough to fit the label. In contrast, the small rectangles of height $2\varepsilon$ might (i) overlap and (ii) be too small for any label. Therefore we have to adapt the algorithm as follows. Firstly, anytime a point $v_j$ on $\ell_v$ is assigned to a point $p_i$ (which implies the placement of a label since the port ratios are fixed), we first remove all rectangles, whose contained center point would be contained within the newly placed label. Secondly, if a combination of a candidate point on $\ell_v$ with a point $p_j$ would imply a label that overlaps an already placed label, the badness function returns $\infty$ for this combination (Figures 10b and 10c). ◀

## B  The SL-leader Case

In this section, we investigate the problem using SL-leaders. Specifically, we show that a number of presented results translate directly or with small adaptions from the OR-case.

### B.1  Fixed Candidate and Fixed Order SL-leader Labelings

We begin with the problems in SL-$C^\triangleq O^\triangleq SA^\triangleq$, which have a candidate set, locked order and locked port ratio. Section 4.1, which investigates the corresponding OR-leader variants, in fact never uses the geometry of the leaders. Since we can also in constant time determine the length of an SL-leader for a pair of a feature and a point on $B$, Theorem 12 does directly cover all problems in SL-$C^\triangleq O^\triangleq SA^\triangleq$.

▶ **Observation 21.** *The problems* SL-$C^{\bullet}O^{\bullet}SA^{\bullet}$ *can be solved in* $O(n^2|C|)$ *time.*

## B.2   Leader Length Minimal SL-leader Labelings are Crossing Free for Free Order

We continue with the problem SL-$C^{\bullet}O^{\prime}S_{\equiv}A_{\equiv}^{\bullet}$, whose OR-pendant was discussed in Section 4.2. The missing part is the following lemma, proving that optimal labelings are crossing-free also for SL-leaders.

▶ **Lemma 22.** *Given an instance of a problem variant in* SL-$CO^{\prime}S_{\equiv}A_{\equiv}$ *any leader-length minimal labeling $L$ is crossing-free.*

**Proof.** Assume a leader-length minimal labeling $L$ contains two crossing leaders $\gamma_1$ and $\gamma_2$ connecting $p_1$ to its port $\xi_L(p_1) = \xi_1$ and $p_2$ to its port $\xi_L(p_2) = \xi_2$, respectively. Let $\gamma_1'$ and $\gamma_2'$ be the two SL-leaders connecting $p_1$ to $\xi_2$ and $p_2$ to $\xi_1$ respectively. Since the leaders cross, $\gamma_1$ and $\gamma_2$ are the diagonals of the quadrilateral $p_1\xi_2\xi_1p_2$. Therefore $|\gamma_1| + |\gamma_2| > |\gamma_1'| + |\gamma_2'|$ and replacing $\gamma_1$ and $\gamma_2$ with $\gamma_1'$ and $\gamma_2'$ results in a labeling with less total leader length than $L$ leading to a contradiction. ◀

Since any pair of feature point and port defines one unique leader, Lemma 22 is sufficient to use the same matching algorithm as in Theorem 16, yielding the following observation.

▶ **Observation 23.** *The problems in* SL-$C^{\bullet}O^{\prime}S_{\equiv}A_{\equiv}^{\bullet}$ *can be solved in* $O(n|C|^2)$ *time.*

## B.3   A Remark about Modelling SL-Leaders Lengths as a Function

When comparing OR-$C^{\prime}O^{\bullet}SA$ to SL-$C^{\prime}O^{\bullet}SA$, a similar approach may initially seems suitable for the problems in SL-$C^{\prime}O^{\bullet}SA$ and it may possibly even seem easier due to the fact that a single feature-port combination implies only a single SL-leader this setting present some obstacles. The placement of the port for $p_1$ still fixes the entire solution and the equivalent of Lemma 18 also holds for SL-leaders, i.e., there are at most $O(n^2)$ continuous connected intervals in which the implied labeling is crossing free. Even when setting up the function for the length of a single leader (equivalent to $g$), which yields $g_{SL}(p_i, \theta) = \sqrt{1 + r^2 + 2r\cos(\theta)}$, we observe that this function has only a single local minimum within the range $0 \le \theta < 2\pi$. However, for the sum $h_{SL}(\theta) = \sum_{i=1}^{n} g_{SL}(p_i, \theta)$ it is not obvious how to bound the number of local minima of the sum, even when assuming that we can efficiently find the local minima of such a function precisely. It is trivial to construct an instance that contains exactly $n$ local minima and in experimental exploration we never encountered an instance that exhibits more than $n$, leading to the following conjecture.

▶ **Conjecture 24.** *The problems in* SL-$C^{\prime}O^{\bullet}SA$ *can be solved exactly in* $O(n^2)$ *time in the real-RAM model of computation.*

## B.4   NP-**Hardness for SL-Leaders**

The hardness result of Section 5 also translates to SL-leaders. However, while we can reuse most of the reduction, some parts require a more detailed look.

▶ **Theorem 25.** *Given an instance of* SL-$C^{\prime}O^{\prime}S_{\equiv}A^{\bullet}$ *together with $k \in \mathbb{R}$ it is (weakly)* NP-*hard to decide if there exists a labeling $L$ with a total leader length of less than $k$.*

**Proof.** We reuse the entire construction as described in the proof of Theorem 20 (and we also reuse the notation). The only change is the definition of the value $k$. For every point $p$ let $d(p)$ be distance of $p$ to $B$. We set $k = \sum_{i=1}^{n} d(p_i) + d(p_U) + d(p_D) + 1$. If we now consider the shortest and longest possible leader for every point in the small circle of radius $\frac{1}{4\pi n}$, we obtain a maximal total difference over all $n$ points of $\frac{1}{2\pi} < 1$.

Now if there exists a partition of $X$ into two sets $X_1, X_2$, s.t., $\sum_{x \in X_1} x = \sum_{x \in X_2} x$, there is a labeling $L$ where the leaders of $p_D$ and $p_U$ are orthogonal to $B$ (and therefore the shortest possible leaders). If in any labeling these two leaders are not both the shortest possible leaders, the smallest angle between them is at most $\pi - \frac{2\pi}{S+2}$ (this is again a direct consequence of the weights being integer). Clearly the sum over their length is minimized if this angle is maximal, and we therefore express their length as shown in Equation 2.

$$\sqrt{2\left(R_1^2 + R_2^2 - R_1 R_2 \cos\left(a \frac{2\pi}{S+2}\right) - R_1 R_2 \cos\left((1-a)\frac{2\pi}{S+2}\right)\right)} \tag{2}$$

Recall that the distance of $p_D$ and $p_U$ to the center of $D$ is $R_1 = \frac{S+2}{4\pi}$ and the radius of $D$ is $R_2 = \frac{S+2}{2\pi}$. For any value of $S$, this expression is minimal for $a = 0.5$ and we restrict our analysis to this case. For any value of $S$ larger than 20, this expression is larger than 1. Therefore if the two leaders are not both the shortest possible leader, the sum of their length is at least 1 larger than their possible minimum. Now let $L$ be a labeling in which the two leaders of $p_U$ and $p_D$ are as short as possible and compare this to a labeling $L'$ in which at least one of the two leaders is not the shortest possible leader. The total leader length of $L$ is at most $k$ even if we assume that all leaders of the points $p_1, \ldots, p_n$ are as long as possible, while the total leader length in $L'$ will always exceed $k$ even if all leader of the inner points are as short as possible. This concludes the reduction. ◀

## C Free Port Ratios

So far we have only investigated problem variants, in which the port ratio for every label has been fixed as part of the input. In this section, we consider the set of problems, where the port ratio can be chosen at runtime, giving us more freedom to optimize the leader routing. It turns out that (partially with some adaptions), a number of solutions of the previous sections can be adapted to these settings. An overview of the relevant results is shown in Table 2.

### C.1 Problem Equivalence for Problems with Uniformly Spaced Ports

We start by providing a lemma, proving the equivalence between two sets of problems, which makes the results for the first set directly applicable to the second. For a problem instance $I$ let $L(I)$ be the leader-length minimal labeling and let $L_\alpha(I)$ be a labeling which is obtained by rotating all labels in a labeling $L(I)$ around $B$ uniformly by an angle of $\alpha$.

▶ **Lemma 26.** *For any problem instance $I$ of $T\text{-}C^x O^y S_\equiv A_\equiv^{\mathscr{P}}$ with $x, y \in \{\mathscr{P}, \text{🔒}\}$ and $T \in \{\mathsf{SL}, \mathsf{OR}\}$, there exists a problem instance $I'$ of $T\text{-}C^x O^y S_\equiv A_\equiv^{\text{🔒}}$ and a value $\alpha$, s.t., the leader-length minimal labeling $L(I)$ is equal to $L_\alpha(I')$.*

**Proof.** Note that for any $x, y$ and $T$ the ports in a labeling of $T\text{-}C^x O^y S_\equiv A_\equiv$ (which covers all problem instances in the lemma statement) are equally distributed around $B$ and all labels have the same size. Therefore in any labeling $L$ for such a problem, the angle between the port $\xi_L(p)$ and start of the label $s_L(p)$ is the same for all feature $p$. By rotating counter-clockwise by this angle (which is also visualized in Figure 11) we obtain a labeling in which $\xi_L(p)$ and

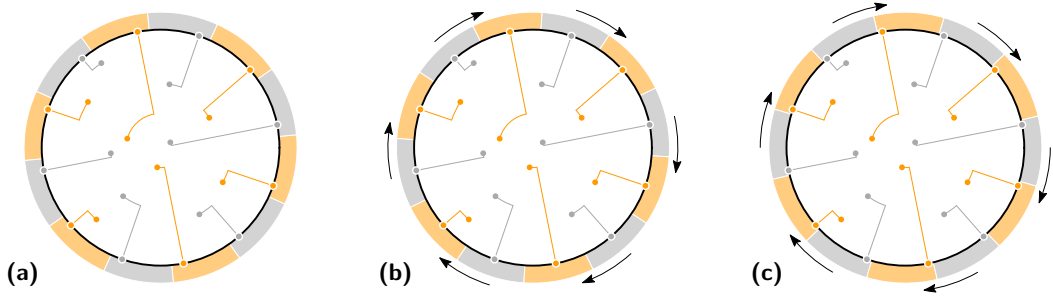

**Figure 11** Any solution with uniform label sizes and a uniform ratio (a) can be rotated (b) to obtain a solution of any other ratio (c).

$s_L(p)$ coincide; note that by our definitions this is not a valid labeling, since $\xi_L(p)$ and $s_L(p)$ necessarily have to have a non-zero distance between them. Nevertheless we will use this as a canonical labeling. Note that the rotation does not change any leader or port placement and therefore does not change the total leader length. To obtain any labeling $L$ from any other labeling $L'$ with a different port ratio, we simply transform both into the canonical labeling and then reverse the second rotation to obtain $L'$ from the canonical labeling. ◂

A direct consequence of the previous lemma is that for any problem instance of a problem in $COS_\equiv A_\equiv^{\text{key}}$ we can simply fix the port ratio to 0.5 to obtain the corresponding problem in $COS_\equiv A_\equiv^{\text{lock}}$ and by the process described in the proof of Lemma 26, we obtain an optimal labeling for the original problem.

▶ **Observation 27.** *Any problem in $T\text{-}C^x O^y S_\equiv A_\equiv^{\text{key}}$ with $x, y \in \{\text{key}, \text{lock}\}$ and $T \in \{\mathsf{SL}, \mathsf{OR}\}$ can be reduced to $T\text{-}C^x O^y S_\equiv A_\equiv^{\text{lock}}$ by choosing a fixed port ratio.*

## C.2 Using Boundary Labeling without Fixed Port Ratio

The algorithm of Benkert et al. [7] labels a set of rectangles to the right of a vertical line. The exact points where the *po*-leaders connect to these rectangles (which corresponds to what we refer to as port ratio) are not fixed in the algorithm as presented by Benkert et al.; recall that we pointed out that a restriction to a specific range can be encoded into the arbitrary badness function. If we do not restrict the function to this specific range, we can instead allow the leader to connect anywhere within a rectangle, which corresponds to a free port ratio. However, since the dynamic program repeatedly subdivides its instance into two separate independent sub-instances these choices of where a single leader connects are made independently for every leader and we can not guarantee that the connection points all collectively correspond to a uniform port ratio.

While we can obtain a set of possible positions for a port of $p_{\min}$ in the same fashion as for a locked port ratio, the reductions we described in Lemma 7, Observation 8 and Lemma 9 can not be used immediately, because without a locked port ratio, the placement of a port of $p_{\min}$ does not fix the position of its label and therefore it also does not fix where we have to place the rectangles on the vertical line in the instance of Boundary Labeling.

We therefore provide lemmas, which show that for any one fixed position of $p_{\min}$, there are at most a certain number of different possible label placements we have to consider. We start with the problems in which we have fixed port candidates and uniform labels.

▶ **Lemma 28.** *For any problem in* OR-$C^\bullet OS_\equiv A_\equiv^\bullet$, *given a port $\xi$ for the $p_{min}$, there are at most $|C|$ possible placements for the label of $p_{min}$ in a leader-length minimal labeling $L$. We can find all possible $|C|$ placements in time $O(|C|)$.*

**Proof.** Assume first that we can not rotate the labels in $L$ clockwise around $B$, without moving a port outside of the range in which it is allowed to connect to its label. In other words, there is a port that has exactly a distance of $\varepsilon$ to the start of its label. Note that if we knew the portion of the port that has this property, this would fix the position of every single other label. Since there are only $|C|$ possibilities for the position of any port, this induces a set of $|C|$ possible labelings. Now assume that no port has this property. Then we could rotate the labels in $L$ freely clockwise around $B$ until one port has this property. Note that rotating just the labels does not change the position of any leader or port and therefore the newly obtained labeling is also leader-length minimal. Since in one possible labeling all labels are defined by a constant offset from a candidate in $C$, we can easily obtain this set in time $O(|C|)$. ◀

The previous proof relied on the fact that the labels are of uniform size. If this is not the case we can still rotate a leader-length minimal labeling clockwise until there is a port with the necessary property. Then, if we are guaranteed that they are in a predefined order, we can consider all $n$ possible labels, which are connected to one candidate in $C$. Since the order of the labels is fixed, the combination of a candidate with the desired property and a given label fixes the position of all other labels and we arrive at the following observation.

▶ **Observation 29.** *For any problem in* OR-$C^\bullet OS_\equiv A_\equiv^\bullet$, *given a port $\xi_1$ for $p_{min}$, there are at most $|C|n$ possible placements for the label of $p_{min}$ in a leader-length minimal labeling $L$. We can find all possible $|C|n$ placements in time $O(|C|n)$.*

We now turn to the setting in which we do not have any predefined port candidates. Observe that for free and non-uniform port ratios, a single port can freely move along $B$ independently of all other ports and – provided that it is still placed in the valid area within its label and the rotation did not introduce any crossings – we still obtain a valid labeling. To control this flexibility we first state the following observation.

▶ **Observation 30.** *Let $L$ be a leader-length minimal labeling for a problem in* OR-$C^\bullet OSA_\equiv^\bullet$. *Then no port of a single feature can be moved clockwise or counter clockwise along $B$, s.t., the connected leader becomes shorter without introducing a crossing between leaders.*

With this observation, we are prepared to state the lemma corresponding to Lemma 28 for the setting without candidate position.

▶ **Lemma 31.** *For any problem in* OR-$C^\circ OS_\equiv A_\equiv^\bullet$, *given a port $\xi$ for $p_{min}$, there are at most $n$ possible placements for the label of $p_{min}$ in a leader-length minimal labeling $L$. We can find all possible $n$ placements in time $O(n)$.*

**Proof.** We again start (analogue to the proof of Lemma 28) by assuming that there is a port in $L$, which is as far counter-clockwise in its label as it can be, i.e., we can not rotate the labels clockwise without invalidating the labeling. Let $\xi'$ be this port. If a movement of $\xi'$ in clockwise or counterclockwise direction along $B$ is impossible, because it would introduce a crossing between two leaders, then $\xi'$ coincides with an intersection of $B$ and a ray starting at $X$ through a feature (of which there are only $n$). If the rotation in both directions is possible then by Observation 30 the leader connected to $\xi'$ has to increase in length in both cases. This implies again that $\xi'$ coincides with an intersection of $B$ and a ray starting at $X$ through a feature. Therefore for a fixed position there are only $n$ possible options for $\xi'$. Finally note that every choice of $\xi'$ fixes the position of one and therefore all labels, including the label which includes $p_{\min}$. Since the labels are uniform and indistinguishable, there are at most $n$ possible placements of the label of $p_{\min}$ we have to consider and we can clearly compute these in $O(n)$ time. ◀

At this point we highlight that we can – similar to the case with candidate positions – leverage a fixed order of the labels to deal with non-uniform sizes, by simply considering all possible $n$ labels for the one that includes $\xi'$ adding an additional factor of $n$.

▶ **Observation 32.** *For any problem in* OR-$C^{\!\!\!\nearrow}O^{\bullet}S_{\equiv}A_{\equiv}^{\bullet}$, *given a port $\xi$ for $p_{min}$, there are at most $n^2$ possible placements for the label of $p_{min}$ in a leader-length minimal labeling $L$. We can find all possible $n^2$ placements in time $O(n^2)$.*

It remains to include the consideration of the different label placements of the label of $p_{\min}$ into the number of different instances of BOUNDARY LABELING we have to create. To do so we state a new version of Theorem 10 which takes this into account.

▶ **Theorem 33.** *For any problem $P$ in* OR-$CO^{\bullet}SA_{\equiv}^{\!\!\!\nearrow}$ *or* OR-$CO^{\!\!\!\nearrow}S_{\equiv}A_{\equiv}^{\!\!\!\nearrow}$ *let $A_P$ and $C_P$ be the sizes of the set of possible ports for $p_{min}$ and the size of the set of possible placements of the label of $p_{min}$, respectively, s.t., the sets can be computed in $O(A_P)$ time and $O(C_P)$ time respectively. Let the time required to solve the created instance of* BOUNDARY LABELING *be $O(B_P)$. Then $P$ can be solved in time $O(A_P \cdot C_P \cdot B_P)$.*

**Proof.** We compute the set of possible ports for $p_{\min}$ as stated in Observation 4, Lemma 5 or Observation 6; recall that neither of these results required locked port ratios and are therefore also applicable. Then, for a given location for the port of $p_{\min}$, we compute the set of possible locations of the label of $p_{\min}$ using Lemma 28, Observation 29, Lemma 31 or Observation 32. The position of this label fixes the position of all other labels since in the relevant problem variants either the order is locked or the label size is uniform and therefore the labels are indistinguishable (solely by size). This gives $A_P \cdot C_P$ different combinations of a port for $p_{\min}$ and its label position. For any such combination, we create an instance of BOUNDARY LABELING and solve it as described in Lemma 7, Observation 8 and Lemma 9; it is important to note that while these results rely on a fixed port position, they only do so to derive the exact position of the label of $p_{\min}$, which we have now also fixed. We obtain a *po*-labeling minimizing the badness function, which has a one-to-one correspondence to a leader length minimal OR-labeling of our original instance. ◀

## C.3  Free Candidates, Free Order and Non-uniform labels remain NP-**hard**

Finally we show that the NP-hardness result of Sections 5 and B.4 extend to free but uniform port ratios, i.e., the problems in $C^{\!\!\!\nearrow}O^{\!\!\!\nearrow}S_{\equiv}A_{\equiv}^{\!\!\!\nearrow}$. Note that the construction of the NP-hardness proofs relied on the label of the two points $p_U$ and $p_D$ partitioning $B$ into two equally sized free pieces. If the port ratio is free, we have no direct control over the exact placement of the labels of these points. However since their labels have the same size, a necessarily equal port ratio also guarantees this property. Further note that for any partition of labels to the left and right side, if there is a labeling with a fixed port ratio for every feature $p_1, \ldots, p_n$ then actually any combination of port ratios – and therefore also specifically the one where all port ratios are uniform – leads to a leader-crossing-free labeling.

▶ **Observation 34.** *Given an instance of $C^{\bullet}O^{\bullet}S_{\doteq}A^{\bullet}_{\doteq}$ together with $k \in \mathbb{R}$ it is (weakly) NP-hard to decide if there exists a labeling $L$ with a total leader length of less than $k$.*

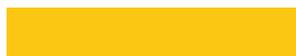